\documentclass[12pt]{article}

\usepackage{amsthm,amssymb,amsmath}
\usepackage{graphicx}
\usepackage{epsfig}

\newcommand{\ggl}{\gamma\gamma}
\newcommand{\gzl}{\gamma \mathrm{Z}}
\newcommand{\ccgg}{\chi\chi\to\gamma\gamma}
\newcommand{\ccgz}{\chi\chi\to\gamma Z}

\newcommand{\gev}{\mathrm{\,\,GeV}}
\newcommand{\sgn}{\mathrm{sgn}}
\newcommand{\svunit}{\mathrm{cm^3s^{-1}}}

\begin{document}
\begin{titlepage}
\centering
\textbf{\huge Gamma ray lines: what will they tell us about SUSY?}\\[1cm]
{\large Carlos E. Yaguna}\\[3mm]
{\small Departamento de F\'{i}sica Te\'orica C-XI, Universidad Aut\'onoma de Madrid, Cantoblanco, E-28049 Madrid, Spain\\ 
Instituto de F\'{i}sica Te\'orica UAM-CSIC, Universidad Aut\'onoma de Madrid, Cantoblanco, E-28049 Madrid, Spain}

\begin{abstract}
Neutralino dark matter can be indirectly detected by observing the gamma ray lines from the annihilation processes $\ccgg$ and $\ccgz$. In this paper we study the implications that the  observation of these two lines could have for the determination of the supersymmetric parameter space. Within the minimal supergravity framework, we find that, independently of the dark matter distribution in the Galaxy, such observations by themselves would allow to differentiate between the coannihilation region, the funnel region, and the focus point region. As a result, several restrictions on the msugra parameters can be derived. Within a more general MSSM scenario, we show that the observation of $\gamma$-ray lines might be used  to discriminate between a bino-, a wino-, and a higgsino-like neutralino, with important consequences for cosmology and for models of supersymmetry breaking. The detection of the $\ggl$ and $\gzl$ lines, therefore, will not only provide an unmistakable signature of dark matter, it will also open a new road toward the determination of  supersymmetric parameters. 
\end{abstract}

\end{titlepage}
\section{Introduction}
The identification of  dark matter is one of the most pressing problems in astroparticle physics today. We know for sure that dark matter exists --it accounts for $23\%$ of the energy-density of the Universe-- but we do not know what it is made of. An appealing framework to explain the dark matter is that of Weakly Interacting Massive Particles --WIMPs. In fact, particles with weak-strength interactions and masses of order of hundreds of $\gev$s generally have the correct properties to account for the dark matter. They behave as cold dark matter and typically yield a  relic density that is not far from the observed dark matter density. Besides, WIMPs can be produced and studied in accelerators such as the LHC as well as be detected, directly and indirectly, in different kinds of dark matter experiments. Hence, WIMPs  provide a   testable and well-motivated  framework to explain the nature of dark matter.

Among the several WIMP dark matter candidates considered in the literature, the lightest neutralino from supersymmetric models have received the most attention. Low energy supersymmetry is indeed the best motivated scenario for physics beyond the standard model. It explains the hierarchy problem, achieves the unification of the gauge couplings, and contains a suitable dark matter candidate: the lightest neutralino.

Neutralinos from the halo can annihilate with each other into standard model particles, providing a means to \emph{indirectly} detect the existence of neutralino dark matter. By searching for the neutralino annihilation products --mainly for gamma rays, positrons, antiprotons, and neutrinos-- one may hope to detect a component due to dark matter annihilations, or at least to be able to  constrain the supersymmetric  parameter space. Several experiments, including PAMELA \cite{pamela} and Fermi \cite{fermi}, are already looking for such signals.  

A common problem with some of these indirect detection searches is that the annihilation signal is pretty much featureless, and the background  is not so well-known. As a consequence, even if an excess over the expected background is observed, it is not easy to attribute it to dark matter rather than to a different background. In fact, we have already witnessed two different realizations of this particular situation: the excesses reported by EGRET and by PAMELA. Several years ago,  the EGRET collaboration observed an excess (over the expected background) in the $\gamma$ ray flux from the Galactic center \cite{Hunger:1997we}.  Such excess could be explained by annihilating dark matter \cite{Bergstrom:1997fj,Cesarini:2003nr} but also by an alternative background model --the so-called \emph{optimized} background \cite{Strong:2004de}. Recently, the Pamela collaboration reported an excess in the positron fraction above $10 \gev$ \cite{Adriani:2008zr}. In some models, such excess can be  explained by annihilating or decaying dark matter \cite{Meade:2009iu}, but an astrophysical explanation, which entails a modified background \cite{Choi:2009qc}, is also viable \cite{Hooper:2008kg,Profumo:2008ms}, and perhaps even favored. From these two situations we learn that it is highly  non-trivial to claim the detection of dark matter in indirect detection experiments.

To unmistakably detect dark matter, one would likely need to combine several indirect detection channels as well as to make use of additional input from  accelerator searches and direct detection experiments. Or, one could look for a background-free signature of dark matter.

The $\gamma$-ray lines produced by neutralino annihilation provide such a signal. They have no plausible astrophysical background and therefore constitute a \emph{smoking gun} signature of dark matter. Their drawback is that, because the processes $\ccgg$ and $\ccgz$ occur at the one-loop level,  the signal is  typically suppressed with respect to other detection channels, and the detection prospects are not as good. In sections  \ref{sec:msugra} and \ref{sec:mssm} we will see that, depending on a number of astrophysical, cosmological, and supersymmetric assumptions, the flux from the $\ggl$ and $\gzl$ lines  can vary over many orders of magnitude. It is fair to say, though, that it typically lies below the 5-year Fermi sensitivity \cite{Baltz:2008wd}. That is, present experiments are not that sensitive to a line signal from neutralino annihilation. For the most part, we will essentially ignore that fact. The relevant point for us is that these lines will eventually be observed, hopefully in the next generation of satellite experiments. What we want to study in this paper are the implications of such observations. In particular, we would like to know if it is possible, at least in principle, to extract, from the observation of the $\gamma$-ray lines, some relevant information about the supersymmetric model.

The observation of a single $\gamma$-ray line, say the $\ggl$ line, will be an unmistakable signature of dark matter, and will further provides us with the value of the neutralino mass, $E_{\ggl}=m_\chi$. The associated gamma ray flux depends, however, on an unknown astrophysical factor, the line-of-sight integral,  that make it difficult to extract from the measurement the value of $\sigma(\ccgg)$, or of any other supersymmetric parameter. Only by assuming a specific halo profile could such values be obtained. But, once both lines have been observed, it is possible to obtain directly from the data the ratio $\sigma(\ccgg)/\sigma(\ccgz)$, a quantity that depends \emph{only} on supersymmetric parameters. That is, it is feasible to use the two line observations to get rid of the astrophysical uncertainty and learned something about the supersymmetric model itself. This fact is not new, it was recognized long time ago. In 1997, Bergstrom et al \cite{Bergstrom:1997fj} explicitly stated that  ``If both lines were to be observed, a comparison of line strengths would give interesting information on the supersymmetric model". They never said, though, what exactly is that interesting information that can be obtained once the two lines are observed. And that question has remained unanswered, or rather unasked, over all these years. This paper tries to give an answer to that question.

To keep our analysis rather general, we study two different supersymmetric scenarios. First, we consider the minimal supergravity model --msugra--, the most well-known framework for supersymmetry breaking. We will see that within msugra the observation of the $\gamma$-ray lines allows us to single out the correct region of the viable parameter space, giving strong restrictions on supersymmetric parameters. Second, we consider a somewhat generic MSSM model defined at low energies. We will  show that in this case it is possible to determine the dominant composition of the neutralino --that is, whether it is bino-like, wino-like or higgsino-like-- from the observation of the $\gamma$-ray lines. Such information might have profound implications for cosmology, indicating a non-standard scenario,  as well as for particle physics, promoting some specific scenarios of supersymmetry breaking.

It must be stressed that our study is only theoretical. That is, we do not consider any particular experiment, nor take into account the expected uncertainties in the observation of the gamma ray lines. Instead, we follow a matter of principle approach:  We just want to know what could be \emph{in principle} learned from such observations. We do hope, however, that our results serve as a strong motivator to search for and detect the gamma ray lines from neutralino annihilation.

The rest of the  paper is organized as follows: In the following section we will present a general discussion about  the gamma ray lines  in supersymmetric models. Section \ref{sec:msugra} is dedicated to the  analysis of the two lines in the msugra model. In section \ref{sec:mssm}, such analysis is extended to a more generic supersymmetric model.  Finally, in section \ref{sec:con} our conclusions are presented.

\section{General considerations}
\label{sec:general}

\begin{figure}[t]
\centering
\begin{tabular}{cccc}
\includegraphics[scale=0.5]{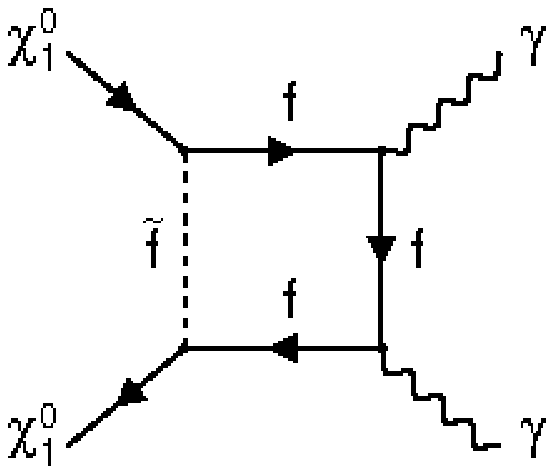} &
\includegraphics[scale=0.5]{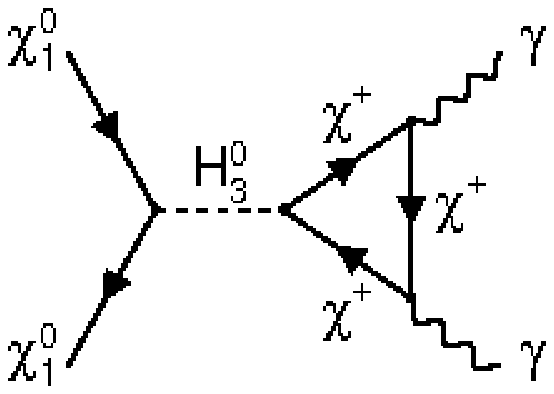} &
\includegraphics[scale=0.5]{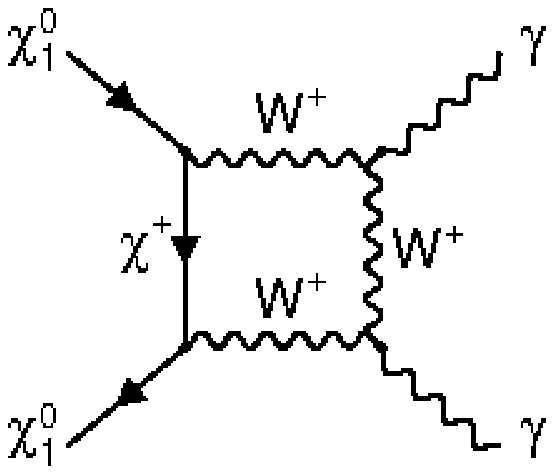} &
\includegraphics[scale=0.5]{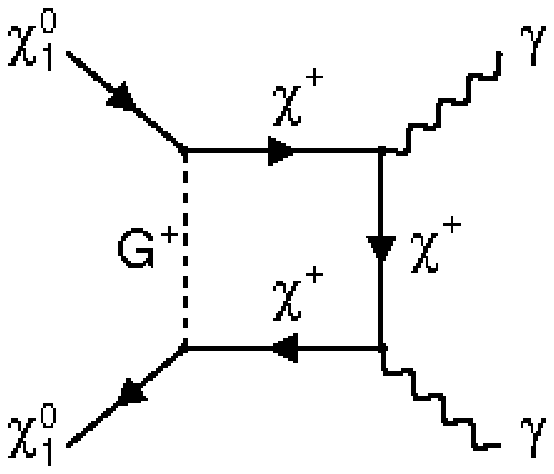}
\end{tabular}
\caption{\small Some of the diagrams that contribute to the process $\ccgg$ in the MSSM.
\label{fig:feynman}}
\end{figure}

The process $\ccgg$ takes place at one-loop level and, for neutralinos from the halo, gives rise to a photon with energy equal to the mass of the neutralino,
\begin{equation}
E_{\ggl}=m_\chi\,.
\label{eq:egl}
\end{equation}
The total  $\ccgg$ amplitude, calculated in \cite{Bergstrom:1997fh}, receives contributions from   four different classes of diagrams:
\begin{equation}
\mathcal{A}=\mathcal{A}_{f\bar f}+\mathcal{A}_{H^+}+\mathcal{A}_W+\mathcal{A}_G\,,
\end{equation} 
where the indices label the particles in the internal loop. They correspond respectively to fermions and sfermions, higgs bosons and charginos, $W$ and charginos, and charginos and Goldstone bosons. Some of the diagrams contributing to this process, one from each class, are shown in figure \ref{fig:feynman}. The resulting amplitudes are lengthy and   complicated expressions that depend non-trivially  on various combinations of MSSM parameters, and are therefore  better studied numerically.

Neutralino annihilation into a photon and a $Z$ boson also occurs at one-loop and gives rise to an almost monochromatic photon of energy
\begin{equation}
E_{\gzl}=m_\chi-\frac{m_Z^2}{4m_\chi}\,.
\label{eq:ezl}
\end{equation}
Thus, $E_{\ggl}>E_{\gzl}$. Again, the total amplitude, computed in \cite{Ullio:1997ke}, receives contributions from four different classes of diagrams,  and the final expression is even more complicated than that for $\ccgg$. In DarkSUSY \cite{Gondolo:2004sc} the full expressions for the annihilation cross sections into $\ggl$ and $\gzl$ --in the limit of vanishing relative velocity of the neutralino pair-- have been implemented. We will use those results throughout this paper.   

The expected gamma ray flux from neutralino annihilation into $\ggl$ or $\gzl$ in the galactic halo is given by
\begin{equation} 
\Phi_\mathrm{\ggl,\gzl}(\psi)=\frac{N_\gamma \sigma v}{8\pi m_\chi^2}\int_\mathrm{line\,of\,sight} \rho^2(l)dl(\psi)
\label{eq:phi}
\end{equation}
where $\psi$ is the angle between the direction of observation and that of the galactic center. In this expression, $\sigma$ is the corresponding cross section and $N_\gamma$ denotes the number of photons in the final state, so $N_\gamma=2$ for $\ccgg$ whereas $N_\gamma=1$ for $\ccgz$. Notice that this flux not only depends on particle physics parameters, such as the neutralino mass and the cross section, but also on astrophysical factors, such as the local dark matter density and the halo profile, through the line of sight integral.

The uncertainty with the halo model is particularly troublesome because, depending on the region of the sky, it could amount to a variation in the flux by several orders of magnitude. Such variation might prevent us from extracting, for an observed flux, direct information on supersymmetric parameters through the cross section. To be able to learn about the supersymmetric model from such observations, two approaches can be considered. One is to observe a $\gamma$-ray line from a region of the sky where the dependence of the line of sight integral with the halo profile is small. An example of such a region is the Galactic centered annulus considered in \cite{Baltz:2008wd}. In that case, the cross section could be obtained from the flux with only a mild uncertainty due to the dark matter distribution. The other possibility is to use the  observation of both lines ($\ggl$ and $\gzl$) to eliminate the dependence with the halo profile. In fact, if the two lines were observed from the same region of the sky, we could take  the ratio between the  fluxes to obtain 
\begin{equation}
\frac{\Phi_{\ggl}}{\Phi_{\gzl}}=\frac{2\sigma(\ccgg)}{\sigma(\ccgz)},
\label{eq:ratio}
\end{equation}
a quantity that no longer depends on unknown astrophysical factors. Could this ratio tell us something relevant about the supersymmetric model? 

In the next sections we will scan the parameter space of two supersymmetric scenarios and, following the two approaches mentioned above, we will study the predictions for the line cross sections and for their ratio.  It will be shown that they indeed provide important information on the fundamental supersymmetric parameters.

\section{$\ggl$ and $\gzl$ lines in msugra}
\label{sec:msugra}

In minimal supergravity models, the soft breaking Lagrangian of the MSSM is determined by only five parameters. They are
\begin{equation}
M_{1/2},m_0,A_0,\tan\beta,\sgn(\mu)\,.
\end{equation}
The first three correspond, respectively, to the universal gaugino mass, the common scalar mass and the trilinear coupling at the GUT scale. $\tan\beta$ is the ratio between the vevs of the two Higgs doublets and $\sgn(\mu)$ is simply the sign of the $\mu$ parameter --its magnitude being determined by the electroweak symmetry breaking conditions. To obtain the soft breaking Lagrangian at the weak scale, its parameters must be run down from the GUT scale to $m_Z$ using the renormalization group equations and the msugra boundary conditions.

The phenomenology of msugra models has been extensively studied in the literature --see e.g. \cite{Baer:2004qq}. Regarding dark matter, the most remarkable feature of msugra models is that the relic density constraint, $\Omega_\chi\sim \Omega_{DM}$, can be satisfied only within narrow regions of the  parameter space. These viable regions are:
\begin{enumerate}
\item The coannihilation region. In this region the neutralino is almost degenerate with the lightest stau ($m_\chi\sim m_{\tilde\tau}$) and coannihilation effects help reduce the relic density. 

\item The funnel region. In this region the neutralino mass is about half the mass of the CP-odd Higgs ($2m_\chi\sim m_A$) and  neutralinos can annihilate efficiently through the $A$-resonance. This region opens up only for large $\tan\beta$ ($\gtrsim 45$) and it is strongly constrained by $b\to s\gamma$. 

\item The focus point region. In this region the $\mu$ parameter is small and the neutralino acquires a non-negligible higgsino component. This component as well as coannihilation effects with higgsino-like charginos and neutralinos  lead to an enhanced annihilation rate. Throughout this region a very large value of $m_0$ is required.
\end{enumerate} 

To study in detail the predictions for the two lines and for their ratio within msugra, we implemented separate scans of the viable parameter space, one for each of the three regions. For simplicity we set $A_0=0$ in all the scans and vary all the  relevant parameters --$M_{1/2}, m_0, \tan\beta, \sgn(\mu)$-- within their allowed range. The scan of the funnel region was obtained imposing $\sgn(\mu)=-1$ and $\tan\beta\geq 40$ from the very beginning. For the focus point region, we used the weak scale parametrization  proposed in \cite{Baer:2005ky}. In all cases we checked that accelerator and phenomenological  constraints were satisfied and that the neutralino relic density, computed with DarkSUSY, was compatible with the observed dark matter density. Our sample consists of $1000$ viable models for each region.   

\begin{figure}[t]
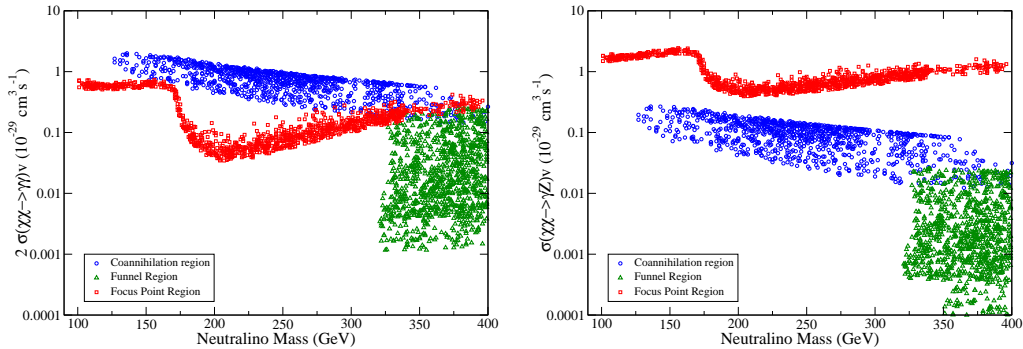

\centering
\begin{tabular}{cc}
\includegraphics[scale=0.25]{ggsugra.eps} &
\includegraphics[scale=0.25]{gzsugra.eps}
\end{tabular}
\caption{\small The cross sections $\sigma(\ccgg)$ (left) and $\sigma(\ccgz)$ (right) as a function of the neutralino mass for the three viable regions of msugra.}
\label{fig:msugra}
\end{figure}

Notice that as the neutralino mass increases, it becomes more difficult to separate the two lines, for they both converge toward $m_\chi$ --see equations (\ref{eq:egl}) and (\ref{eq:ezl}). It is therefore unrealistic to assume that both lines can be observed irrespective of the neutralino mass. For any given experiment (or a given energy resolution) there will always be a maximum neutralino mass beyond which the two lines are not separable, and only the sum $\Phi_{\ggl}+\Phi_{\gzl}$ can be observed. If the relative energy resolution is $\epsilon=\sigma_E/E$, the two lines are separable only if
\begin{equation}
m_\chi\lesssim \frac{M_Z}{\sqrt{4\epsilon}}\,.
\end{equation}
For this reason, in the following we will only consider neutralino masses below $400\gev$. They give rise to separable lines provided that $\epsilon\gtrsim 0.013$.

\begin{figure}[t]
\centering
\includegraphics[scale=0.4]{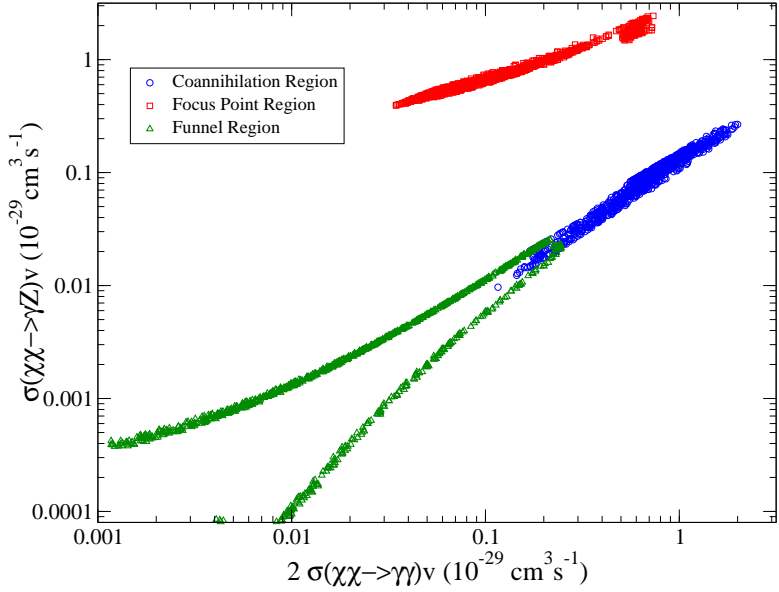}
\caption{\small The cross sections for the $\ggl$ and $\gzl$ lines in the three viable regions of msugra.\label{fig:gggzsugra}}
\end{figure}

The cross sections for $\ccgg$ and $\ccgz$ are shown  in figure \ref{fig:msugra} as a function of the neutralino mass for the three viable regions of msugra. Notice that, due to the $b\to s\gamma$  constraint, points within the funnel region feature neutralino masses larger than about $300 \gev$. For the other two regions, neutralino masses span essentially the whole range $100-400 \gev$.  In that range,  the total variation in the cross sections is about three and four orders  of magnitude, respectively for $\ccgg$ and  $\ccgz$. Within a given region, however, the variation is smaller. In the coannihilation region, for instance,  $\sigma(\ccgg)$ and $\sigma(\ccgz)$ vary only by about one order of magnitude. From the figure we see that the largest value of $\sigma(\ccgg)$ is obtained, over most of the mass range, within the coannihilation region. Only in the high mass range, $m_\chi\gtrsim 350 \gev$, can the focus point region (and even the funnel region) provide with comparable values. The largest value of  $\sigma(\ccgz)$, on the other hand, always lies within the focus point region. It is also clear from the figure that the behavior of these cross sections with respect to the neutralino mass depends on the region considered. They decrease with the neutralino mass along the coannihilation region but increase with it over most of the focus point region. In fact, it only decreases in a narrow region around the top quark mass. That feature can be easily explained: once the annihilation channel into $t\bar t$ gets open, the annihilation rate increases and, consequently, the required higgsino component in the lightest neutralino gets reduced. And it is precisely that component what mainly determines the cross sections for neutralinos within the focus point region.

Figure \ref{fig:gggzsugra} omits the mass dependence and instead shows the three regions in the plane $\sigma(\ccgg)$ vs $\sigma(\ccgz)$. Notice that the two branches of the funnel region, corresponding to $2m_\chi>m_A$ and $2m_\chi<m_A$, are clearly differentiated in this plane. Because  the focus point region is widely separated from the other two, and the coannihilation and funnel regions only intersect in a narrow band, it is possible to pinpoint the correct viable region  from the knowledge of the two cross sections. Within the coannihilation and the focus point region, it is also possible to predict the value of one of the cross sections from  the other one, once the correct viable region has been identified.

\begin{figure}[t]
\centering
\includegraphics[scale=0.4]{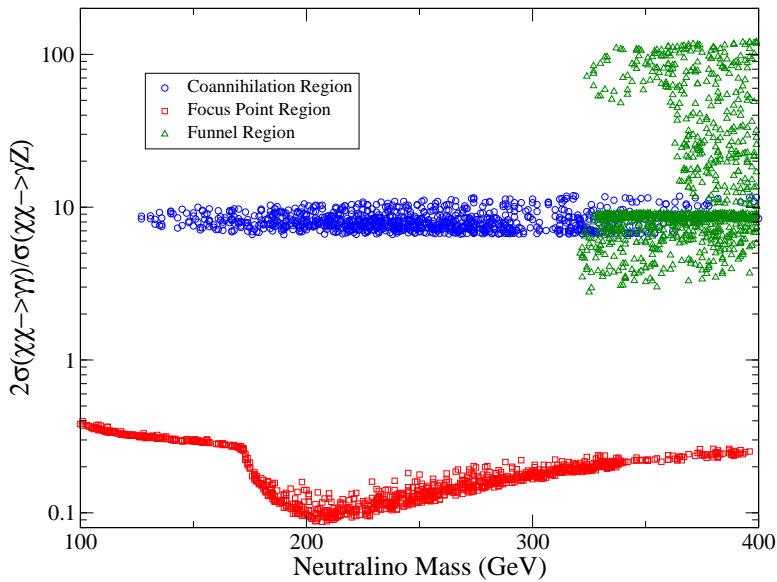}
\caption{\small The ratio of the cross sections for the $\ggl$ and $\gzl$ lines as a function of the neutralino mass in the three viable regions of msugra.\label{fig:corrsugra}}
\end{figure}

Figure \ref{fig:corrsugra} shows the ratio of the two line cross sections as a function of the neutralino mass for the three viable regions of msugra. For the focus point region, the ratio lies between $0.1$ and $0.4$, that is $\sigma(\ccgz)$ is between $2.5$ and $10$ times larger than $2\sigma(\ccgg)$.  Along the coannihilation region, the ratio $2\sigma(\ccgg)/\sigma(\ccgz)$ lies in a narrow band around $10$, approximately between $7$ and $12$. Hence, $2\sigma(\ccgg)$ is typically $10$ times larger than $\sigma(\ccgz)$. In the funnel region, on the contrary, the ratio spans a wide range between $3$ and $100$. That is, $2\sigma(\ccgg)$ could be up to two orders of magnitude larger than $\sigma(\ccgz)$. 

This figure illustrates one of the main results of this paper. By combining the measurements of the two lines, it is possible to obtain direct information on the underlying supersymmetric model, independently of the dark matter distribution.  Within msugra, for instance, a ratio smaller than $1$ would unambiguously point toward the focus point region. Analogously, a ratio larger than $10$ (or between $3$ and $7$) would indicate a model within the funnel region. If, on the other hand, the ratio is found to be around $10$, two possible solution exists, one in the coannihilation region and one in the funnel region. In that case, we could use the neutralino mass, which is deduced from the position of the lines, as a further discriminant. A neutralino mass below $300 \gev$, for example, would clearly single out the coannihilation region as the correct solution. 

\begin{figure}[t]
\centering
\includegraphics[scale=0.4]{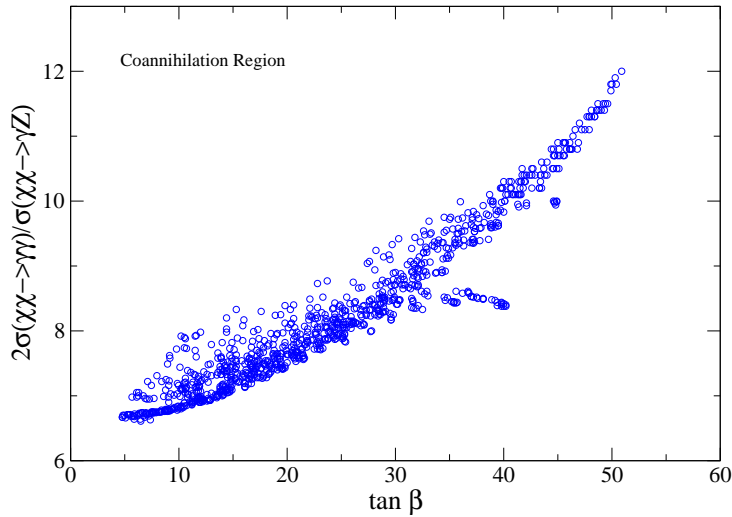}
\caption{\small The ratio of the cross sections for the $\ggl$ and $\gzl$ lines as a function of $\tan\beta$ for points lying in the coannihilation region of msugra.\label{fig:cotbsugra}}
\end{figure}

It is important to stress that the identification of the  viable region provides a lot of additional data on  supersymmetric parameters. Once we know that the focus point region is realized, for instance, we know that $m_0$ is necessarily very large and, consequently, the sfermion spectrum should be heavy. If, instead, it is the funnel region that has been established, then  $\tan\beta$ must be large and  $m_A$ should be about twice the neutralino mass. Similarly, within the coannihilation region one certainly expects an scalar tau almost degenerate with the neutralino.

One may wonder whether any additional constraints on msugra parameters, besides the viable region and its implications, could be extracted from the ratio $2\sigma(\ccgg)/\sigma(\ccgz)$. After studying its behavior with respect to supersymmetric parameters along the three viable regions, we have found only one more useful piece of information that could be derived from it: the value of $\tan\beta$ along the coannihilation region. Within the coannihilation region the ratio $2\sigma(\ccgg)/\sigma(\ccgz)$ increases with $\tan\beta$ --see figure \ref{fig:cotbsugra}. It goes from about $7$ for $\tan\beta=5$ to about $12$ for $\tan\beta=50$. Hence, if this ratio were measured one could possibly say if $\tan\beta$ is large or small.

Summarizing, we have seen that in msugra models  the observation of the gamma ray lines from neutralino annihilation will provide us with a lot of useful details about the underlying supersymmetric model. Withouth any additional input from accelerator searches or other dark matter detection experiments and independently of the dark matter distribution in the Galaxy, such observations could easily select, among the viable regions of msugra, the one that is actually realized in nature. They would also provide valuable information on supersymmetric parameters such as $m_\chi$, $m_0$, $\tan\beta$, and $\mu$.  

\section{$\ggl$ and $\gzl$ lines in the MSSM}
\label{sec:mssm}

Even if well-motivated, msugra is just one of the many possibilities for supersymmetry breaking that have been considered in the literature. In this section, we will generalize the study of the $\ggl$ and $\gzl$ lines and their implications in two important ways: We will consider a more generic supersymmetric spectrum and we will not impose the dark matter constraint.

The supersymmetric spectrum could be very different from that predicted in the msugra scenario. In fact, all different kinds of non-universality --in gauginos as well as  in the higgs and the scalar sector-- have been considered and analyzed in the literature (see for example \cite{Belanger:2004ag}). Regarding the dark matter phenomenology, the most important feature of these models is that, in constrast with msugra, the lightest neutralino does not have to be a  bino-like neutralino.  In anomaly mediated supersymmetry breaking models, for instance, gaugino masses are not universal at the GUT scale and the dark matter candidate turns out to be a wino-like neutralino \cite{Randall:1998uk}. Other non-universal models based on $SU(5)$ may also give rise to a higgsino-like lightest neutralino \cite{Chattopadhyay:2003yk}. To take this possibilities into account, we will consider  rather generic supersymmetric models defined at the weak scale by the following set of parameters:
\begin{equation}
M_1, M_2, M_3, \mu, m_A, m_{\tilde f}, \tan\beta\,.
\end{equation}
They correspond respectively to the three gaugino masses, the $\mu$ parameter, the mass of the CP-odd higgs, the common soft-breaking scalar mass, and the ratio between the vevs of the two Higgs doublets, $\tan\beta$.  In our scan, we allow all mass parameters to take values up to $2$ TeV and require $\tan\beta<60$. This set of parameters is large and rich enough to allow for significant deviations from the msugra spectrum.

\begin{figure}[t]
\centering
\includegraphics[scale=0.4]{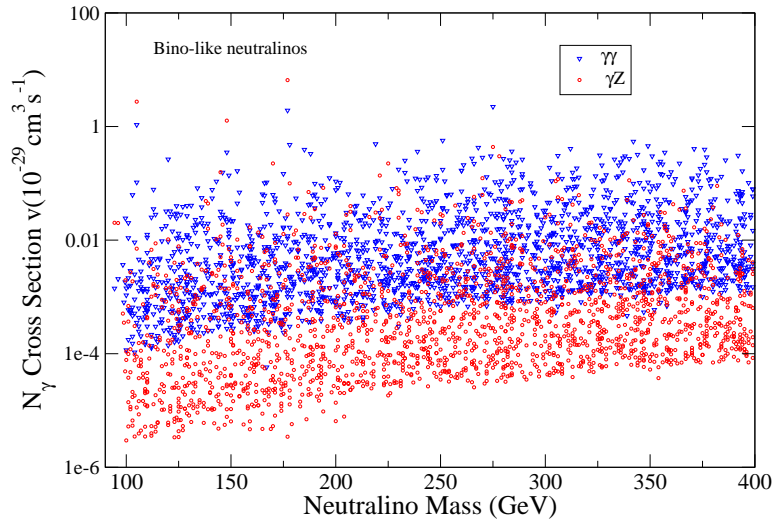}
\caption{\small The cross sections for the $\ggl$ line and  $\gzl$ line for bino-like neutralinos as a function of the neutralino mass.\label{fig:binomneu}}
\end{figure}

The relic density constraint plays an ambiguous role in dark matter phenomenology. On the one hand, it is very effective in reducing the viable parameter space. It typically is the most stringent constraint on supersymmetric models. On the other hand, it is well-known that it can be easily avoided \cite{Kamionkowski:1990ni,Gelmini:2006pq}. In fact, the computation of the relic density \emph{assumes} that the standard cosmology was valid during the time of neutralino freeze-out ($T\gtrsim 10 \gev$), but there is no direct evidence that the universe was radiation dominated before Big-Bang Nucleosynthesys ($T\sim 1$ MeV). In non-standard cosmological scenarios, the history of the universe before BBN differs from the usual radiation dominated scenario and, as a result, the predicted dark matter density might take another value. Typically, the neutralino relic density depends on additional cosmological parameters that can be adjusted so as to obtained the observed dark matter density \cite{Gelmini:2006pq}. That is why, it makes sense not to impose the usual dark matter constraint within such models.

For dark matter detection purposes, this non-standard scenarios are more than welcome, as they allow for neutralino annihilation rates much larger than those found for a standard thermal relic. Light winos and higgsinos, for instance, are perfectly valid dark matter candidates within these scenarios. Given this possibility, in this section we will impose phenomenological and accelerator constraints but not the dark matter constraint.

Our sample of models is divided into three different sets, one for bino-like neutralinos, one for wino-like neutralinos, and one for higgsino-like neutralinos. In all cases, we require the dominant neutralino composition to be larger than $90\%$. Hence, in the bino set the lightest neutralino has a  bino component  larger than $90\%$, whereas in the higgsino set it is the higgsino component that is larger than $90\%$. Effectively, this generates models featuring rather pure binos, winos, or higgsinos.   It must be stressed that in supersymmetric models, pure neutralinos tend to be the norm rather than the exception. Mixed neutralinos are also possible but they typically require some degree of fine-tuning between supersymmetric parameters. Our sample consists of $2000$ models for each set. 

\begin{figure}[t]
\centering
\includegraphics[scale=0.4]{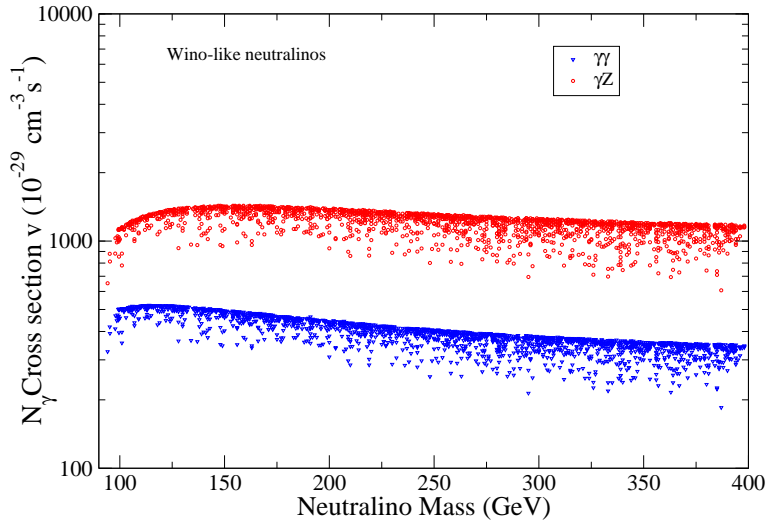}
\caption{\small The cross sections for the $\ggl$ line and  $\gzl$ line for wino-like neutralinos as a function of the neutralino mass.\label{fig:winomneu}}
\end{figure}

The cross section for the processes $\ccgg$ and $\ccgz$ are shown in figure \ref{fig:binomneu} as a function of the neutralino mass for bino-like neutralinos.  Notice that both cross cross section are small, $<10^{-29} \svunit$, and span a wide range, reaching $10^{-33}\svunit$ for $\ccgg$ and even smaller values for $\ccgz$. This large variation in the cross section is mainly due to the scalar mass, with larger  scalar masses correlated with  smaller cross sections.

For wino-like neutralinos  the two cross sections are shown  as a function of the neutralino mass in figure \ref{fig:winomneu}. They are both much larger than those for  bino-like neutralinos. The $\gamma Z$ cross section is around $10^{-26}\svunit$ and has a very narrow range of variation. The $\gamma\gamma$ cross section is smaller, but not by much. It lies between $5\times 10^{-27}$ and $2\times 10^{-27}$. Notice that, according to \cite{Baltz:2008wd}, some of these cross sections are within the expected sensitivity of Fermi-LAT.

\begin{figure}[t]
\centering
\includegraphics[scale=0.4]{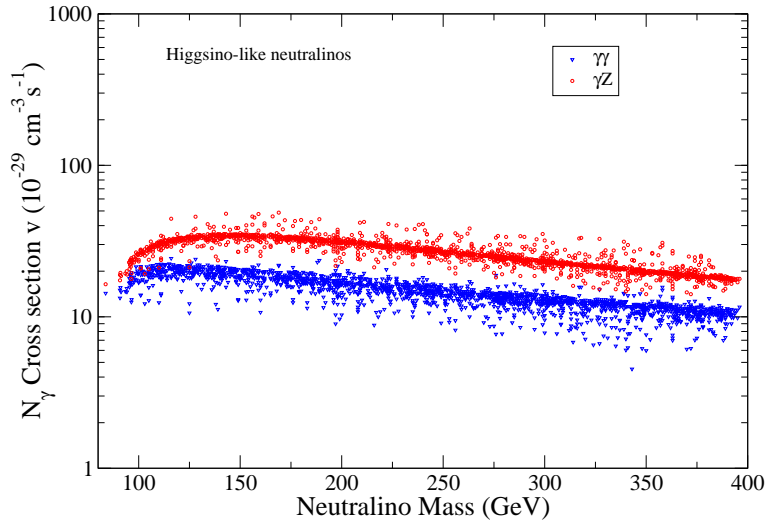}
\caption{\small The cross sections for the $\ggl$ line and  $\gzl$ line for higgsino-like neutralinos as a function of the neutralino mass.\label{fig:hinomneu}}
\end{figure}

Higgsino-like neutralinos feature cross sections into monochromatic photons of order $10^{-28} \svunit$, as illustrated in figure \ref{fig:hinomneu}. As was the case with winos, the cross section into $\gamma Z$ turns out to be always larger, by a factor of two or so, than that into $\gamma\gamma$. And the range of variation of the cross sections is equally small. Hence, for wino- and higgsino-like neutralinos, the value of the cross sections is essentially determined by the neutralino mass.

\begin{figure}[t]
\centering
\includegraphics[scale=0.4]{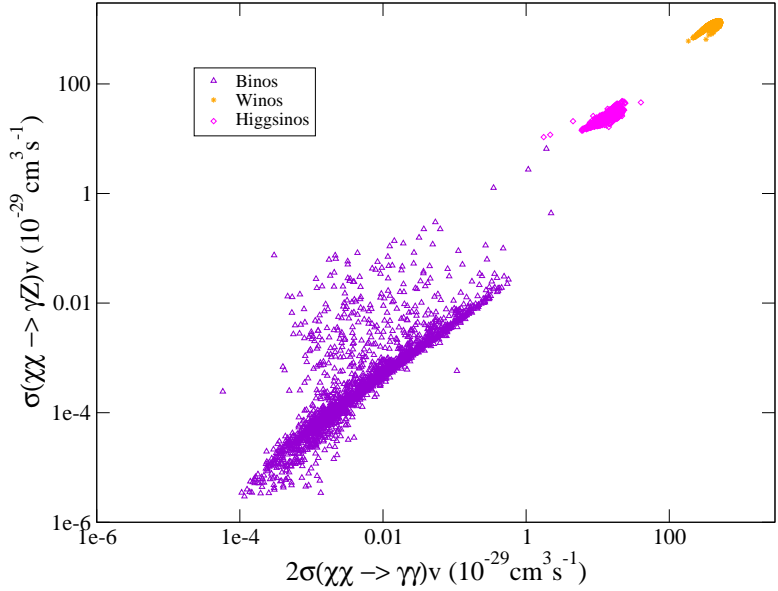}
\caption{\small The cross sections for the $\ggl$ line versus the cross section for  $\gzl$ line for bino-, wino-, and higgsino-like neutralinos in the MSSM. \label{fig:gggz}}
\end{figure}

	Figure \ref{fig:gggz} shows the locations of the three sets of supersymmetric models in the plane $\sigma(\ccgg)$ vs $\sigma(\ccgz)$. It is clear from the figure that by far the largest cross sections are obtained for wino-like neutralinos. The next-to-largest ones correspond to higgsino-like neutralinos and the smallest ones to bino-like neutralinos. Moreover, models featuring wino- and higgsino-like neutralinos occupy very small volumes in the plane. Thus, if one of the two cross sections were known, one could easily predict, using the figure,  the value of the other. For bino-like neutralinos this procedure is not feasible, as there is a lot of dispersion in the cross sections.  Finally, notice also that it is possible to identify the dominant composition of the neutralino from the value of any of the two line cross sections. Thus, $\sigma v\lesssim 10^{-29}\svunit$ corresponds to bino-like neutralinos, $10^{-29}\svunit <\sigma v<10^{-27}\svunit$ is consistent with higgsino-like neutralinos, and $\sigma v>10^{-27}\svunit$ for wino-like neutralinos.

\begin{figure}[t]
\centering
\includegraphics[scale=0.4]{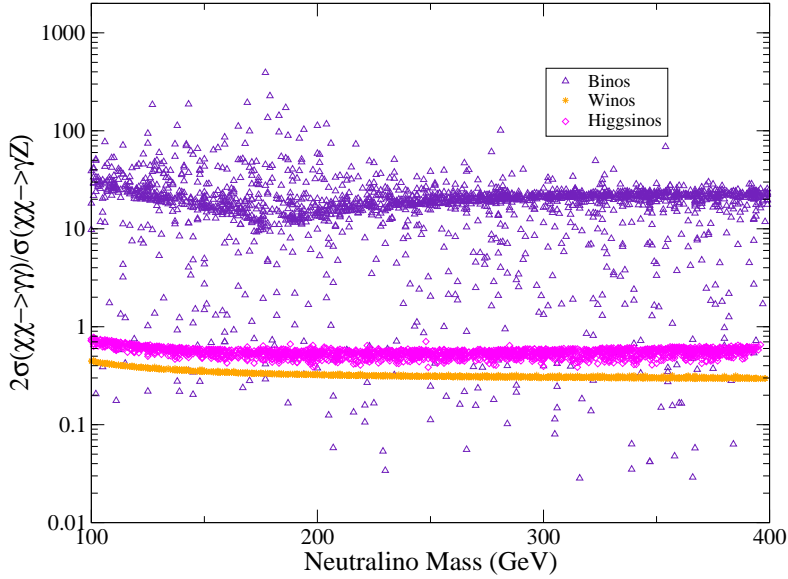}
\caption{\small The cross sections for the $\ggl$ line versus the cross section for the   $\gzl$ line for bino-, wino-, and higgsino-like neutralinos as a function of the neutralino mass.\label{fig:corr}}
\end{figure}

The ratio between the two cross sections as a function of the neutralino mass is illustrated in figure \ref{fig:corr}. Notice that the dependence with the neutralino mass is very weak for the three sets. Models with wino-like neutralinos are all concentrated along a narrow band between $0.5$ and $0.3$. 
Models with higgsino-like neutralinos occupy a slightly wider area between $0.8$ and $0.4$. Models with bino-like neutralinos, on the other hand, feature a large spread in the ratio. Most of them  are concentrated in a wide region between $40$ and $10$, but several models lie well above and well below that area. Ratios smaller than $0.1$ as well as larger than $100$ can be found among them. For that reason, bino-like neutralinos make it more difficult to determine the dominant nature of the lightest neutralino from the observation of the two lines. It is not impossible, though. If that ratio is found to be larger than $1$, for instance, then we can be  certain that the neutralino is bino-like. If the ratio is found within the higgsino band then the neutralino is likely to be higgsino-like, though it could also be bino-like. A similar conclusion would follow if the ratio is found within the wino band. Since the wino and the higgsino regions do not overlap, it should be possible to discriminate between them.

The identification of the dominant neutralino composition might have  profound and far reaching implications. If, for example, a light wino- or higgsino-like neutralino is single out from the observation of the $\gamma$-ray  lines, we would know that gaugino masses do not unify at the GUT scale and, more importantly, that a non-standard cosmological scenario is in play.

\section{Conclusions}
\label{sec:con}

We have studied the possible implications that the observation of the $\ggl$ and $\gzl$ lines from neutralino annihilation will have for the determination of the supersymmetric parameter space. In msugra models, we showed that, independently of the halo profile, such observations may allow to single out the correct viable region for neutralino dark matter. Such identification in turn implies strong restrictions on supersymmetric parameters. We also studied a more generic MSSM model defined at the weak scale and embeded within a non-standard cosmological framework. In such scenarios, we found that it is possible to use the $\gamma$-ray line measurements for discriminating between a bino-like, a wino-like, and a higgsino-like neutralino. Thus, the observation of the $\gamma$-ray lines from neutralino annihilation not only will provide unmistakable evidence of dark matter, it might also give some clues about the cosmological model, and it will certainly restrict the parameter space of supersymmetric models.

\section*{Acknowledgments}
I am supported by the \emph{Juan de la Cierva} program of the Ministerio de Educacion y Ciencia of Spain. I acknowledge additional support from  Proyecto Nacional FPA2009-08958,  the ENTApP Network of the ILIAS project RII3-CT-2004-506222, and the Universet Network MRTN-CT-2006-035863.


\begin{thebibliography}{99}


\bibitem{pamela}
http://pamela.roma2.infn.it/index.php

\bibitem{fermi}
http://fermi.gsfc.nasa.gov/


\bibitem{Hunger:1997we}
  S.~D.~Hunter {\it et al.},
  Astrophys.\ J.\  {\bf 481} (1997) 205.



\bibitem{Bergstrom:1997fj}
  L.~Bergstrom, P.~Ullio and J.~H.~Buckley,
  Astropart.\ Phys.\  {\bf 9}, 137 (1998)
  [arXiv:astro-ph/9712318].

\bibitem{Cesarini:2003nr}
  A.~Cesarini, F.~Fucito, A.~Lionetto, A.~Morselli and P.~Ullio,
  Astropart.\ Phys.\  {\bf 21} (2004) 267
  [arXiv:astro-ph/0305075].

\bibitem{Strong:2004de}
  A.~W.~Strong, I.~V.~Moskalenko and O.~Reimer,
  Astrophys.\ J.\  {\bf 613} (2004) 962
  [arXiv:astro-ph/0406254].



\bibitem{Adriani:2008zr}
  O.~Adriani {\it et al.}  [PAMELA Collaboration],
  Nature {\bf 458} (2009) 607
  [arXiv:0810.4995 [astro-ph]].

\bibitem{Meade:2009iu}
  P.~Meade, M.~Papucci, A.~Strumia and T.~Volansky,
  arXiv:0905.0480 [hep-ph]. And the references therein.


\bibitem{Choi:2009qc}
  K.~Y.~Choi and C.~E.~Yaguna,
  arXiv:0906.0736 [hep-ph].


\bibitem{Hooper:2008kg}
  D.~Hooper, P.~Blasi and P.~D.~Serpico,
  JCAP {\bf 0901}, 025 (2009)
  [arXiv:0810.1527 [astro-ph]].

\bibitem{Profumo:2008ms}
  S.~Profumo,
  arXiv:0812.4457 [astro-ph].

\bibitem{Baltz:2008wd}
  E.~A.~Baltz {\it et al.},
  JCAP {\bf 0807} (2008) 013
  [arXiv:0806.2911 [astro-ph]].



\bibitem{Bergstrom:1997fh}
  L.~Bergstrom and P.~Ullio,
  Nucl.\ Phys.\  B {\bf 504} (1997) 27
  [arXiv:hep-ph/9706232].


\bibitem{Ullio:1997ke}
  P.~Ullio and L.~Bergstrom,
  Phys.\ Rev.\  D {\bf 57} (1998) 1962
  [arXiv:hep-ph/9707333].

\bibitem{Gondolo:2004sc}
  P.~Gondolo, J.~Edsjo, P.~Ullio, L.~Bergstrom, M.~Schelke and E.~A.~Baltz,
  JCAP {\bf 0407} (2004) 008
  [arXiv:astro-ph/0406204].


\bibitem{Baer:2004qq}
  H.~Baer, A.~Belyaev, T.~Krupovnickas and J.~O'Farrill,
  JCAP {\bf 0408} (2004) 005
  [arXiv:hep-ph/0405210].





\bibitem{Baer:2005ky}
  H.~Baer, T.~Krupovnickas, S.~Profumo and P.~Ullio,
  JHEP {\bf 0510} (2005) 020
  [arXiv:hep-ph/0507282].


\bibitem{Belanger:2004ag}
  G.~Belanger, F.~Boudjema, A.~Cottrant, A.~Pukhov and A.~Semenov,
  Nucl.\ Phys.\  B {\bf 706} (2005) 411
  [arXiv:hep-ph/0407218].
  D.~G.~Cerdeno and C.~Munoz,
  JHEP {\bf 0410} (2004) 015
  [arXiv:hep-ph/0405057].
  J.~R.~Ellis, T.~Falk, K.~A.~Olive and Y.~Santoso,
  Nucl.\ Phys.\  B {\bf 652} (2003) 259
  [arXiv:hep-ph/0210205].
  V.~Berezinsky, A.~Bottino, J.~R.~Ellis, N.~Fornengo, G.~Mignola and S.~Scopel,
  Astropart.\ Phys.\  {\bf 5} (1996) 1
  [arXiv:hep-ph/9508249].


\bibitem{Randall:1998uk}
  L.~Randall and R.~Sundrum,
  Nucl.\ Phys.\  B {\bf 557} (1999) 79
  [arXiv:hep-th/9810155].
  G.~F.~Giudice, M.~A.~Luty, H.~Murayama and R.~Rattazzi,
  JHEP {\bf 9812} (1998) 027
  [arXiv:hep-ph/9810442].
  P.~Ullio,
  JHEP {\bf 0106} (2001) 053
  [arXiv:hep-ph/0105052].


\bibitem{Chattopadhyay:2003yk}
  U.~Chattopadhyay and D.~P.~Roy,
  Phys.\ Rev.\  D {\bf 68} (2003) 033010
  [arXiv:hep-ph/0304108].


\bibitem{Kamionkowski:1990ni}
  M.~Kamionkowski and M.~S.~Turner,
  Phys.\ Rev.\  D {\bf 42} (1990) 3310.
  T.~Moroi and L.~Randall,
  Nucl.\ Phys.\  B {\bf 570} (2000) 455
  [arXiv:hep-ph/9906527].
  S.~Profumo and P.~Ullio,
  JCAP {\bf 0311} (2003) 006
  [arXiv:hep-ph/0309220].


\bibitem{Gelmini:2006pq}
  G.~Gelmini, P.~Gondolo, A.~Soldatenko and C.~E.~Yaguna,
  Phys.\ Rev.\  D {\bf 74} (2006) 083514
  [arXiv:hep-ph/0605016].



\end{thebibliography}
\end{document}